\documentclass[12pt]{article}
\usepackage{graphicx}
\input epsf
\usepackage{amssymb}
\parskip=\medskipamount
   \def\CL{{\cal L}}

   \def\Ga{\Gamma}

\def\te{\theta}   
\def\om{\omega}   
\def\ID{\relax{\rm l\kern-.18 em D}}
\def\IE{\relax{\rm l\kern-.18 em E}}
\def\IK{\relax{\rm l\kern-.18 em K}}
\def\IL{\relax{\rm I\kern-.18 em L}}
\def\IN{\relax{\rm I\kern-.18 em N}}
\def\IR{\relax{\rm I\kern-.18 em R}}
\def\R{{\mathbb{R}}}      
\def\uno{\relax{\rm 1\kern-.18 em l}}

\def\IK{\relax{\rm l\kern-.18 em K}}
\def\IL{\relax{\rm I\kern-.18 em L}}
\def\IN{{\Bbb N}}
\def\IR{{\Bbb  R}}

\def\dfrac#1#2{{\displaystyle\frac{#1}{#2}}}

\def\wt{\widetilde}
\def\frac#1#2{{#1\over #2}}
\def\fracpd#1#2{\frac{\partial #1}{\partial #2}}

\def\ptos{\leaders\hbox to 2mm{\hfil{.}\hfil}\hfill}
\def\\{\hfill\break}

\def\<#1>{\langle#1\rangle}

\def\dddot#1{\stackrel{...}{#1}}

\font\tenfrak=eufm10  \font\sevenfrak=eufm7  \font\fivefrak=eufm5
\newfam\frakfam
\textfont\frakfam=\tenfrak\scriptfont\frakfam=\sevenfrak
\scriptscriptfont\frakfam=\fivefrak

\font\tengoth=eufm10 scaled\magstep1 \font\sevengoth=eufm7
\font\fivegoth=eufm5
\newfam\gothfam
\textfont\gothfam=\tengoth\scriptfont\gothfam=\sevengoth
   \scriptscriptfont\gothfam=\fivegoth

\newtheorem{proposicion}{Proposition}

\def\today{\ifcase\month\or
    January\or February\or March\or April\or May\or June\or
    July\or August\or September\or October\or November\or December\fi
    \space\number\day, \number\year}

\begin{document}

\title{Higher-order  Abel equations:  Lagrangian formalism, \\
 first integrals   and Darboux polynomials }

\author{Jos\'e F. Cari\~nena$^{\dagger}$, Partha Guha$^{\ddagger}$$^{\S}$
and Manuel F. Ra\~nada$^{\dagger}$  \\ [2pt]
$^{\dagger}$
  {\sl Departamento de F\'{\i}sica Te\'orica and IUMA, Facultad de Ciencias} \\
  {\sl Universidad de Zaragoza, 50009 Zaragoza, Spain}  \\ [2pt]
$^{\ddagger}$ {\sl Max Planck Institute for Mathematics in the Sciences}\\
{\sl Inselstrasse 22, D-04103 Leipzig, Germany}          \\
$^{\S}$ {\sl S.N. Bose National Centre for Basic Sciences, JD Block}\\
{\sl  Sector-3, Salt Lake, Calcutta-700098, India} }

\date{October 9, 2009}
\maketitle

\begin{abstract}
A geometric approach is used to study a family of higher-order
nonlinear Abel equations. The inverse problem of the Lagrangian
dynamics is studied in the particular case of the second-order
Abel equation and it is proved the existence of two alternative
Lagrangian formulations, both Lagrangians being of a non-natural
class (neither potential nor kinetic term). These higher-order
Abel equations are studied by means of their Darboux polynomials
and Jacobi multipliers. In all the cases a family of constants of
the motion is explicitly obtained. The general $n$-dimensional
case is also studied.
\end{abstract}

\begin{quote}
Higher-order Abel equations. Lagrangian formalism. Constants of
the motion. Darboux polynomials.  Jacobi multipliers.

{\sl Running title:}{\enskip}
Higher-order  Abel equations.


\end{quote}

\footnoterule
{\noindent\small
$^{a)}${\it E-mail address:} {jfc@unizar.es}  \\
$^{b)}${\it E-mail address:} {partha@bose.res.in}  \\
$^{c)}${\it E-mail address:} {mfran@unizar.es}

\newpage

\section{Introduction}

The first-order Riccati equation
$$
  y' = P(x) y^2 + Q(x) y + R(x)
$$
is important mainly because it is a nonlinear one but directly
related to the general linear differential equation of
second-order via a Cole-Hopf transformation. It is usually
considered as the first instance in the study of nonlinear
equations \cite{Dav62} and is endowed with many interesting
properties. For example, it is a Lie system admitting a nonlinear
superposition principle and it is the only nonlinear equation of
the form $y'=f(x,y)$, where $f(x,y)$ is a rational function of the
variable $y$ with coefficients analytic in $x$, that possesses the
Painlev\'e property (nevertheless the Lie-Scheffers theory or the
Painlev\'e approach will not be considered in this paper).

The (first-order) Riccati equation is therefore a nonlinear
equation that has been intensively studied by many authors.  The
important point is that it has been proved that it admits
higher-order generalisations which  are also studied by making use
of several different approaches \cite{Er76,Er77,GrL99} (according
to Davis these higher-order equations were first considered by
Vessiot in 1895).  All the higher-order Riccati equations can be
linearised via a Cole-Hopf transformation to linear differential
equations. It is known, that the higher-order Riccati equations
play the role of B\"acklund transformations for integrable partial
differential equations of higher-order than the KdV equation. The
Riccati chain without potential is naturally associated to Fa\'a
di Bruno polynomials. The Fa\'a di Bruno polynomials appear in
several branches of mathematics and physics and can be introduced
in several ways.

In fact higher-order Riccati equations are related to the
existence of symmetries \cite{MoLe00,EuEuLe07}, Darboux
polynomials \cite{Da1878, La96, MaPr04} and Jacobi multipliers
\cite{NuLe04,Nu05,NuLe08}. We also mention that the second-order
Riccati equation has been studied in \cite{CRS05Riccati} from a
geometric perspective and it has been proved to admit two
alternative Lagrangian formulations, both Lagrangians being of a
non-natural class (neither potential nor kinetic term). An
analysis of the higher-order Riccati equations and all these
properties (Lagrangians, symmetries, Darboux polynomials and
Jacobi multipliers) is presented in \cite{CGR09}.

The Abel differential equation can be considered as the simplest
nonlinear extension of the Riccati equation \cite{Mur60,
Zwi97,PoZai03}. The Abel equation of the first kind
\cite{BriFra98,  Boy03, ChebRo03, Pan05, Boy06, FrRoYo08} is given by
$$
y^{\prime}=f_0(x)+f_1(x)y+f_2(x)y^2+f_3(x)y^3  \,.
$$
There is also another related equation, called Abel equation of
second kind, given by
$$
[g_0(x)+g_1(x)y]y^{\prime}=f_0(x)+f_1(x)y+f_2(x)y^2+f_3(x)y^3 \,,
$$
which is reducible to the previous one \cite{Sch98, Sch98b} and it
is not going to be considered in this paper. On one hand it has
striking similarities with the Riccati equation but on the other
side, as the non-linearity is of higher degree, the properties are
different (and in fact more difficult to be studied).  The
objective of this paper is to study a chain of  higher-order Abel
equations using as an approach the analysis of the differential
geometric properties, the Lagrangian formalism and the theory of
Darboux polynomials and Jacobi multipliers.

The plan of the article is as follows: In Section 2 we review the
hierarchy of higher-order  Riccati equations and then introduce in
a similar way the hierarchy of higher-order  Abel equations.
Section 3 is devoted to a particular case of second-order Abel
equation. We study the existence of a Lagrangian formulation,
obtain some constants of the motion and establish the relationship
of this equation with the theory of Darboux polynomials and Jacobi
multipliers. Section 4 is devoted to the third and fourth-order
equations and in Section 5 we consider the general $n$-dimensional
case. Finally in Section 6 we make some brief comments.

\section{Riccati and Abel equations}

Let us start with the definition of higher-order  Riccati
equations. It is known that  these equations can be obtained by
reduction from the Matrix Riccati equation. The matrix Riccati
equation plays an important part in the theory of linear
Hamiltonian systems, the calculus of variations, and other related
topics.

\subsection{Hierarchy of higher-order  Riccati equations}

Let us denote by $\ID_R$ the following differential operator,
depending on a real parameter $k\in\mathbb{R}$, that will be
called `differential operator of Riccati'
$$
 \ID_R = \frac{d}{dt} + k\,x(t) \,,
$$
in such a way that the action of $\ID_R$ leads to the following
family of differential expressions
\begin{eqnarray*}
 \ID_R^0  x &=& x\cr
 \ID_R  x  &=& \Bigl(\frac{d}{dt} + k x\Bigr)\,x= \dot{x} + k x^2\cr
 \ID_R^2 x &=& \Bigl(\frac{d}{dt} + k x\Bigr)^2x= \ddot{x} + 3 k x \dot{x} + k^2 x^3\cr
 \ID_R^3 x &=& \Bigl(\frac{d}{dt} + k x\Bigr)^3x= \dddot  x + 4 k x  \ddot{x} + 6 k^2x^2 \dot{x} + 3 k  \dot{x}^2 + k^3 x^4 \cr
 \ID_R^4 x &=& \Bigl(\frac{d}{dt} + k x\Bigr)^4x= x^{iv)} + 5 k x \dddot x + 10 k \dot{x} \ddot{x} + 15k^2 x \dot{x}^2 + 10 k^2 x^2 \ddot{x} + 10 k^3 x^3 \dot{x} + k^4 x^5\,.
\end{eqnarray*}
 The  Riccati equation of order  $m$ of the higher-order  Riccati
hierarchy (o chain),  is given by
$$
  \ID_R^m  x = 0 \,,\quad m = 0,1,2,\dots
$$
In fact, the most general form of a Riccati equation of order $m$
is just a superposition of all the previous equations (linear
combination the different members of the hierarchy)
$$
 (p_0 \ID_R^n + p_1 \ID_R^{n-1} + \dots + p_{n-1} \ID_R +  p_n)x + p_{n+1} = 0 \,,
$$
where each $p_i$ is a function of $t$.

These equations have certain properties that make them interesting
from both  physical and mathematical points of view.
Next we point out some of them.
\begin{enumerate}
\item[(1)] The higher-order Riccati equation of order $m$, member
of the Riccati hierarchy, admits the maximal number of Lie point
symmetries that can admit  an equation of order $m$.

\item[(2)] The higher-order Riccati equation of order $m$ can be
linearised and presented as a linear equation of order
$m+1$.

\item[(3)] The dimensional reduction of a linear equation of order
$m+1$ leads to the Riccati equation of order $m$.
\end{enumerate}

\subsection{Hierarchy of higher-order Abel equations}

The most natural generalisation of the Riccati equation is
$$
 \dot x = f(t,x),
$$
where $f(t,x)$ is  a polynomial in the variable $x$ (with coefficients
depending on $t$). The particular case of $f(t,x)$ being  a cubic
polynomial
$$
 f(t,x) =  A_0(t)+ A_1(t) x + A_2(t) x^2 + A_3(t) x^3 \,,
$$
is called Abel equation. Such an equation can be
considered as the simplest nonlinear extension of the Riccati
equation.

Let us denote by $\ID_A$ the following differential operator,
depending on a real parameter $k\in\mathbb{R}$, to be called
`Abel differential operator',
$$
 \ID_A = \frac{d}{dt} + k\,x^2(t) \,,
$$
in such a way that the action of $\ID_A$ leads to a family of
$k$-dependent differential equations whose first members are given
by
\begin{eqnarray*}
 \ID_A^0  x &=& x\cr
 \ID_A  x &=&  \Bigl(\frac{d}{dt} + k x^2\Bigr)\,x = \dot{x} + k x^3\cr
 \ID_A^2  x &=&  \Bigl(\frac{d}{dt} + k x^2\Bigr)^2x= \ddot{x} + 4 k x^2 \dot{x} + k^2 x^5\cr
 \ID_A^3  x &=&  \Bigl(\frac{d}{dt} + k x^2\Bigr)^3x=  \buildrel{...}\over{x} + 5 k x^2 \ddot{x} +
8 k x \dot{x}^2 + 9 k^2 x^4  \dot{x} + k^3 x^7 \cr
 \ID_A^4  x &=&  \Bigl(\frac{d}{dt} + k x^2\Bigr)^4x =
x^{iv)} + 2 k (4 \dot{x}^3 + 13 x \dot{x}\ddot{x}+ 3 x^2\buildrel{...}\over{x})
 + 2 k^2 x^3 (22 \dot{x}^2 + 7 x\dot{x}) + 16k^3 x^6\dot{x} + k^4 x^9
\end{eqnarray*}
We call  this family the hierarchy of higher-order  Abel
equations. The  Abel equation of order  $m$, written in the
so-called simplified form,  is given by
 $$
  \ID_A^m  x = 0 \,,\quad m = 0,1,2,\dots
$$
Actually, the most general form of the Abel equation of order $m$
is just a superposition of all the previous equations (linear
combination of the different members of the hierarchy with
functions $p_i(t)$ as coefficients)
$$
 (p_0 \ID_A^n + p_1 \ID_A^{n-1} + \dots + p_{n-1} \ID_A +  p_n)x + p_{n+1} = 0 \,.
$$
An important point is that the Abel equation can not be obtained, as the Riccati equation, by a reduction procedure from  a linear equation.

\section{Abel equation of second-order }

In this section we will analyse the particular case of the
second-order Abel equation. In particular, we describe the
Lagrangian formulation of the second-order Abel equation.

\subsection{Lagrangian formalism}

The action of $\ID_A^2$ on the function $x(t)$ leads to the
nonlinear equation
\begin{equation}
 \frac{d^2 x}{dt^2} + 4 k x^2 \Bigl(\frac{d x}{dt}\Bigr) + k^2 x^5  =
0  \,, \label{AbelEq2}
\end{equation}
that represents the Abel equation of second-order. It can be
presented as a system of two first-order equations
$$
\left\{\begin{array}{rcl}
 \dfrac{dx}{d t} &=& v  \cr
\dfrac{dv}{d t} &=& - 4 k x^2  v - k^2 x^5
\end{array}\right.
$$ 
that determines a dynamical system that, in differential geometric
terms, is represented by the following vector field
\begin{equation}
  \Ga^{(2)} = v\,\fracpd{}{x} + F_{A2}\,\fracpd{}{v} \,,\qquad
  F_{A2} = - 4 k x^2  v - k^2 x^5 \,.
\end{equation}
defined on the phase space $\R^2$ with coordinates $(x,v)$.

It has been proved in \cite{CRS05Riccati} that the second-order
Riccati equation
$$
 \frac{d^2 x}{dt^2}  + 3 k x \Bigl(\frac{d x}{dt}\Bigr) + k^2 x^3  = 0
$$
can be considered as the Lagrange equation
determined by the following Lagrangian
$$
  L_R =  \frac{1}{v + k\,x^2}  \,. \label{LagR1}
$$

\begin{proposicion}
The nonlinear  Abel equation of second-order (\ref{AbelEq2})
admits a Lagrangian formulation with a non-polynomial Lagrangian.
\end{proposicion}
{\sl Proof:}   There are two different ways of obtaining a
Lagrangian function for the nonlinear Abel equation;  the
Helmholtz approach and the generalisation of the method used for the
corresponding Riccati
case.

The Helmholtz conditions are a set of conditions that a multiplier
matrix $g_{ij}(x,\dot{x},t)$ must satisfy in order for a given
system of second-order equations
$$
  \ddot{x}_j = f_j(x,\dot{x},t)  \,,\quad j=1,2,\dots,n,
$$
when written of the form
$$
  g_{ij}\ddot{x}_j = g_{ij}f_j(x,\dot{x},t)  \,,\quad i,j=1,2,\dots,n,
$$
to be the set of Euler-Lagrange equations for a certain Lagrangian $L$
\cite{Sar82, CrPrTh84, Lopus99, CaRa99} (the summation  convention on
repeated indices is assumed). If a matrix solution $g_{ij}$ is
obtained then it can be identified with the Hessian matrix of $L$,
that is $g_{ij} = \partial L/\partial v_i\partial v_j$, and a
Lagrangian $L$ can be obtained by direct integration of the
$g_{ij}$ functions. The two first conditions just impose
regularity and symmetry of the matrix $g_{ij}$; the two other ones are
equations introducing relations among the derivatives of $g_{ij}$
and the derivatives of the functions $f_i$. Here we only write the
fourth set of conditions that determine the time-evolution of the
$g_{ij}$
$$
\Ga(g_{ij}) =  g_{ik}A_{kj}+ g_{jk}A_{ki} \,,\quad
A_{ab}=-\frac{1}{2}\fracpd{f_a}{v_b} \,.
$$
When the system is one-dimensional  we have $i=j=k=1$ and then
the three first sets of conditions become trivial and the fourth
one reduces to one single first-order P.D.E.
\begin{equation}
\Ga(g) + \Bigl(\fracpd{f}{v}\Bigr) g \equiv v\Bigl(\fracpd{g}{x}\Bigr) + f
\Bigl(\fracpd{g}{v}\Bigr) + \Bigl(\fracpd{f}{v}\Bigr) g= 0  \,,\label{EqHelm}
\end{equation}
that in the case of the Abel equation becomes
\begin{equation}
  v\Bigl(\fracpd{g}{x}\Bigr) - (4 k x^2 v + k^2 x^5)
  \Bigl(\fracpd{g}{v}\Bigr) - 4 k x^2 g = 0 \,.  \label{EqHelmAbel}
\end{equation}
So, the problem reduces to find the function $g$ as a solution of
this equation. Once a solution $g$  is known a Lagrangian $L$ is
obtained by integrating two times the function $g$. The funtion
$L$ obtained from $g$ is unique up to addition of a gauge term

Next we consider the second method that is specific for this
particular nonlinear problem.   The starting point is the idea
that, since the Abel equation is very close related with the
Riccati equation,  it seems natural to assume that the Abel
Lagrangian must be a non-polynomial  function similar to that of the
second-order Riccati equation.

Let us begin by considering the following one degree of freedom
Lagrangian
\begin{equation}
  L = \frac{1}{(v + k\,U(x,t))^m}  \,.   \label{LagUtm}
\end{equation}
From such a Lagrangian  we arrive to the following second-order
nonlinear equation
\begin{equation}
 \ddot{x} + \Bigl(\frac{2+m}{1+m}\Bigr)  \,k \,U_x  \,\dot{x}
 + \Bigl(\frac{1}{1+m}\Bigr)  \,k^2 U  U_x + k U_t = 0  \,.
 \label{Eq2Ut}\end{equation}
Hence, in the particular case of $U$ and $m$ being given by
$$
 U (x,t)=  x^3 \,,\quad  m=2 \,,
$$
then the Lagrangian (\ref{LagUtm}) leads to (\ref{AbelEq2}).
Thus the second-order Abel equation (\ref{AbelEq2}) turns out to be  the
Euler-Lagrange equation of the Lagrangian function
\begin{equation}
  L_A =  \frac{1}{(v + k\,x^3)^2}  \,.  \label{LagAbel1}
\end{equation}
Finally, as a byproduct of this approach, we have also obtained
the Lagrangians for the whole family of nonlinear equations
(\ref{Eq2Ut}) depending of a function $U$.
$\Box$

As a corollary of this proposition we can state that when the
function $U$ is time-independent the nonlinear equation
(\ref{Eq2Ut}) has a first-integral that can be interpreted as a
preserved energy. That is, if we restrict the study to nonlinear
equations arising from a time-independent Lagrangian of the form
$$
  L =   \frac{1}{(v + k\,U(x))^m}   \label{LagUm}
$$
then we can define an associated Lagrangian energy $E_L$ by the
usual procedure
$$
 E_L = \Delta(L) - L \,,\quad   \Delta = v\,\fracpd{}{v} \,,
$$
and we arrive to
$$
 E_L = \frac{-\,\,((1+m)\,v +\,k\,U(x))}{(v + k\,U(x))^{m+1}}
\,,\qquad \frac{d}{dt}E_L = 0.
$$
In the particular case of the Abel Lagrangian $L_A$ we have
\begin{equation}
 E_{L_A} = -\,\,\frac{(3 v + k\,x^3)}{(v + k\,x^3)^3} \,,\qquad
 {d \over dt}\,E_{L_A} =  0 \,. \label{EL1}
\end{equation}
Note that $L_A$ is non-natural and, as there is neither kinetic
term $T$ nor potential function $V$, the energy cannot be of the
standard form $E_L=T+V$. But, in spite of its rather peculiar
form, $E_{L_A}$ is a conserved function for the Abel equation.

An important property of the Lagrangian formalism is that for one
degree of freedom systems if an equation admits a Lagrangian
formulation then the Lagrangian is not unique \cite{CuS66,HoH81}.
This property can be proved in two different ways.   First, the
Helmholtz equation (\ref{EqHelmAbel}) is a linear equation in
partial derivatives and thus it admits many different particular
solutions. Moreover it is clear from the  form of the equation
(\ref{EqHelm}) that if $g_1$ is a particular solution then
$g_2=fg_1$ with $\Ga(f)=0$ is also a solution. A second method is
related with the properties of the symplectic formalism.  In a
two-dimensional manifold all the symplectic forms must be
proportional. Hence if $\om_L$  is known then any other symplectic
form $\om_2$ must be proportional to $\om_L$, that is $\om_2 =
f\om_L$. Then
$$
 i(\Ga_L)\,\om_2 = f\,i(\Ga_L)\,\om_L = f\,dE_L \,.
$$
The right-hand side is an exact one-form if, and only if,
$df{\wedge}dE_L=0$, which shows that $f$ must be a function of
$E_L$. In this case it can be proved that the new symplectic form
$\om_2$ is derivable from an alternative Lagrangian $L_2{\ne}L$ for
$\Ga_L$.

In the particular case of  the Abel system $\Ga^{(2)}$,  several
alternatives Lagrangians can be obtained that, in most of cases,
are of non-algebraic character (with logarithmic terms).
Nevertheless, in the particular case of $f$ given by  $f =
(-1/E_{L_A})^{4/3}$, we have obtained the following algebraic
function
\begin{equation}
 \wt{L}_A  =  (3 v + k\,x^3)^{2/3}    \label{LagAbel2}
\end{equation}
as a new alternative Lagrangian for the Abel equation
(\ref{AbelEq2}).   This new Lagrangian is equivalent to $L_A$ in
the sense that both determine the same dynamics. It determines a
new energy $\wt{ E}_{L_A}$ that is a constant of the motion for
the Abel equation; nevertheless it must not be considered as a new
fundamental constant  since it is  a function of the original
energy $E_{L_A}$.

\subsection{Constants of the motion and geometric formalism }

A function $T$ that satisfies the following property
$$
 {d \over dt}\,T \ne  0  {\quad},\dots,{\quad} {d^m \over dt^m}\,T \ne 0
 \,,\quad {d^{m+1} \over dt^{m+1}}\,T=0 \,,
$$
is called a generator of integrals of motion of degree $m$. Notice
that this means that the function $T$ is a non-constant function
generating a constant of motion by successive time derivations.

Let us denote by $T_1^{(2)}$ the following function
$$
  T_1^{(2)}  =  \frac{x}{v + k\,x^3} \,.
$$
Then we have that under the evolution given by Abel's equation
(\ref{AbelEq2})
$$
 \frac{d}{d t}\,T_1^{(2)} = T_2^{(2)} =   1 \,,{\quad}
 \frac{d}{dt}\,T_{2}^{(2)} = 0 \,.
$$
Thus, the function $J_{t1}$ defined by
$$
 J_{t1} = T_{1}^{(2)} -  t   \,,
$$
is a time-dependent constant of the motion for the Abel equation.

This means that  we have obtained two constants of the motion (of
quite different nature) for the Abel equation of second-order: the
energy $E_{L_A}$ and the time-dependent function $J_{t1}$.

In differential geometric terms a time-independent Lagrangian
function $L$ determines an exact two-form $\om_L$ defined as
$$
 \te_L = \Bigl(\fracpd{L}{v_x}\Bigr) \,dx  \,,\qquad
 \om_L = -\,d\te_L  \,,
$$
and  $L$ is said to be regular when the 2-form $\om_L$ is
symplectic. In the particular case of $L$ given by
(\ref{LagAbel1}) $\om_{L_A}$ is given by
$$
  \om_{L_A} = \Bigl(\frac{6}{(v + k\,x^3)^4} \Bigr)\,dx\,{\wedge}\,dv \,,
$$
and the dynamical vector field $\Ga^{(2)}$ is the solution of the
equation
$$
 i(\Ga^{(2)})\,\om_{L_A} = dE_{L_A}   \,.
$$

Next we consider two interesting class of symmetries:  `master
symmetries' and `non-Cartan symmetries'.   The idea is that,  in
differential geometric terms, constants of motion that depend of
the time but in a polynomial way are related  with the existence
of master symmetries  \cite{SaC81, Da93,Fe93, Ra99} and in some
very particular cases with non-Cartan symmetries.

Given a  dynamics represented by a certain vector field
$\Ga$,  then a vector field $Z$  satisfying
$$
 [Z,\Ga] \,=\,{\widetilde Z} \ne 0  \,,{\qquad}
 [\,{\widetilde Z}\,,\Ga]  = 0  \,,
$$
is called a `master symmetry' of degree $m=1$ for $\Ga$.
When $Z$ is such that
$$
 [Z,\Ga] = {\widetilde Z} \ne 0  \,,\
 [\,{\widetilde Z}\,,\Ga]   \ne 0  \quad {\rm and}\quad
 [\,[\,{\widetilde Z}\,,\Ga]\,,\Ga] = 0 \,,
$$
then $Z$ is called a `master symmetry' of degree $m=2$.
The generalisation to higher values of $m$ is straightforward:
$$({\rm ad}(\Gamma))^{m+1}(Z)=0,\quad {\rm but}\quad ({\rm
  ad}(\Gamma))^{m}(Z)\ne 0\,.
$$

It is well-known that symmetries are important because they give
rise to constants of the motion and reduction procedures. Master
symmetries, which are a rather peculiar class of symmetries,
determine time-dependent constants of motion (the system is
time-independent but the constant is however time-dependent). This
can be seen as follows: if $Z$ is a master symmetry of degree one,
the time-dependent vector field $Y_Z$ determined by $Z$ as follows
\cite{Ra99}
$$
  Y_Z =  Z + t\, [Z,\Ga] + (\frac{1}{2})\,t^2\,[\,[Z,\Ga]\,,\Ga]
$$
is a time-dependent symmetry of $\Ga_t= \partial/\partial t+\Ga $,
which is the suspension of the vector field $\Gamma$ \cite{AbrMar78}.
This symmetry determines a time-dependent constant of motion $J_t
= T - t\,\Ga(T)$ that depends linearly of $t$ (for $m=2$ the
corresponding constant $J_t$ will be quadratic in $t$ and for
$m=3$ will be cubic).

Let  $Z_1$ be  the Hamiltonian vector field of the
function $T_{1}^{(2)}$, that is, the unique solution of the equation
$$
 i(Z_1)\,\om_L = dT_{1}^{(2)}  \,,
$$
which is given by
$$
 Z_1 = -\,(\frac{1}{6}) P_{A1}^2\,
\Bigl(\, x\fracpd{}{x} +(v - 2 k x^3)\,\fracpd{}{v}\,\Bigr) \,,\qquad
P_{A1} =  v + k\,x^3 \,.
 $$
Then $Z_1$ is
 a symplectic symmetry (that is, ${\CL}_{Z_1}\,\om_L = 0$)
because it is the Hamiltonian vector field of $T_{1}^{(2)}$, and moreover
it is a dynamical symmetry because
$$
i([Z_1\,,\,\Ga^{(2)}]\om_L=i(Z_1)({\mathcal{L}}_{\Ga^{(2)}}\omega_L)-{\mathcal{L}}_{\Ga^{(2)}}(i(Z_1)\om_L)=-
{\mathcal{L}}_{\Ga^{(2)}}  (dT_{1}^{(2)})=0\,,
$$
and therefore, as $\om_L$ is non-degenerate,    $[Z_1\,,\,\Ga^{(2)}] = 0$.

Note however  that $Z_1$ is not a symmetry of the energy since $Z_1(E_{L_A})
\ne 0$.

Thus, $Z_1$ is a dynamical but non-Cartan symmetry of the
Lagrangian system \cite{Cr83, LoMaRa99}. These symmetries are
rather peculiar and only appear in some very particular cases. In
particular it was proved in \cite{LoMaRa99} that if the
Hamiltonian vector field $X_F$ with the function $F$ as
Hamiltonian in a symplectic manifold $(M,\omega)$  is a dynamical
but non-Cartan symmetry, then $X_F(H)$ must be a numerical
constant $X_F(H)=\alpha\ne 0$. In this case we are considering,
$F=T_{1}^{(2)}$ and we have $Z_1(E_{L_A})=\alpha=-1$.

We close this section by recalling that the Riccati equation was
endowed with similar properties  but the function  $T_{1}^{(2)}$
was the Lagrangian $L_R$ itself \cite{CGR09}.

\subsection{Darboux polynomial and Jacobi multiplier approach }

The existence of constants of the motion and the Lagrangian
inverse problem for polynomial vector fields are two questions
related with two important ideas: Jacobi multipliers and Darboux
polynomials.

Let $U$ be an open subset of ${\mathbb R}^n$. We say that a
polynomial function ${\mathcal{D}} : U \to {\mathbb R}$ is a
Darboux polynomial for a polynomial vector field $X$ if there is a
polynomial function $f$ defined in $U$ such that $X{\mathcal{D}}=f
\mathcal{D}$  \cite{Da1878, La96, MaPr04,CGR09}. The function $f$
is said to be the cofactor corresponding to such a Darboux
polynomial and the pair $(f ,\mathcal{D})$ is called a Darboux pair.

When $f  = 0$, then the Darboux  polynomial is a first integral.
We say that $\mathcal{D}$ is a  proper Darboux polynomial if $f\ne
0$. If ${\mathcal{D}}_1$
and  ${\mathcal{D}}_2$ are Darboux polynomials with the same cofactor,
the quotient  ${\mathcal{D}}_1/{\mathcal{D}}_2$ is a first integral.

On the other side given a vector field $X$ in an  oriented
manifold $(M,\Omega)$, a function $R$ such that $R\, i(X)\Omega$
is closed is said to be a Jacobi multiplier (JM) for $X$. Recall that
the divergence of the vector field $X$ (with respect to the volume
form $\Omega$) is defined by the relation
$$
 \mathcal{L}_X\Omega = ({\rm div\,}X)\, \Omega\,.
$$
This means that $R$ is a multiplier  if and only if
 $R\, X$ is a divergenceless vector field and
then
$$
 {\mathcal{L}}_{RX}\Omega = ({\rm div\,}RX)\, \Omega
 = [X(R)+R\, {\rm div}X]\,\Omega=0\,,
$$
and therefore we see that $R$ is a last multiplier for $X$ if and
only if
\begin{equation}
 X(R) + R\,{\rm div}X = 0\,.       \label{Jmcond}
\end{equation}
Note that if $R$ is a never vanishing Jacobi multiplier, then $f
R$ is a Jacobi multiplier too if and only if $f$ is a constant of
motion. We also note that the equation (\ref{EqHelmAbel}) can now be considered 
as a particular case of the equation (\ref{Jmcond}).

The remarkable point is that if ${\mathcal{D}}_1, \ldots,
{\mathcal{D}}_k$, are Darboux polynomials with corresponding
cofactors $f_i$,  $i=1,\ldots,k$, one can look for multiplier
factors of the form
\begin{equation}
 R=\prod_{i=1}^k{\mathcal{D}}_i^{\nu_i}    \label{Relem}
\end{equation}
and then
$$
 \frac{X(R)}{R} =\sum_{i=1}^k\nu_i\frac{X({\mathcal{D}}_i)}{{\mathcal{D}}_i}
 = \sum_{i=1}^k\nu_i\, f_i\,,
$$
and therefore, if the coefficients $\nu_i$ can be chosen such that
\begin{equation}
 \sum_{i=1}^k\nu_{i}\, f_i=-{\rm div\,}X\label{expcond}
\end{equation}
holds, then we arrive to
$$
 \frac{X(R)}{R} =\sum_{i=1}^k\nu_i\,f_i=-{\rm div\,}X\,,
$$
and consequently $R$ is a Jacobi last multiplier for $X$.

Finally, if $R$ is a Jacobi multiplier for a  vector field which
corresponds to a second-order differential equation, there is an
essentially unique Lagrangian $L$ (up to addition of a gauge term)
such that $R=\partial^2L/\partial v^2$ \cite{NuLe04,Nu05,NuLe08}.

From these general concepts  we can return to the Abel equation.
In this case  the polynomial ${\mathcal{D}_1}$ defined  by
$$
 {\mathcal{D}}_1(x,v) = v+kx^3
$$
is a Darboux polynomial for $ \Ga^{(2)}$ with cofactor $-kx^2$ since
$$
 \left(v\,\fracpd{}{x} + F_{A2}\,\fracpd{}{v} \right)(v+kx^3)
 = - kx^2(v+kx^3)\,.
$$
The  divergence of the vector field  $ \Ga^{(2)}$ is $-4kx^2$, and
then, according to (\ref{expcond}), we see that there is a
multiplier of the form
$$
 R= {\mathcal{D}}_1^{\nu_1}\,,
$$
with   $\nu_1=-4$. Consequently, the Abel equation admits a
Lagrangian description by means of a function $L_1$ such that
$$
 \fracpd{^2L_1}{v^2} = (v+kx^3)^{-4}\,,
$$
from where we obtain the Lagrangian $L_1=L_A$ given by (\ref{LagAbel1}).

But  the polynomial ${\mathcal{D}_2}$ defined  by
$$
 {\mathcal{D}}_2(x,v)=3v+kx^3
$$
is a Darboux polynomial for $ \Ga^{(2)}$ with cofactor $-3kx^2$,
because
$$
 \left(v\,\fracpd{}{x} + F_{A2}\,\fracpd{}{v} \right)(3v+kx^3) =
 3kx^2v-3(4kx^2v+k^2x^5) =-3kx^2(3v+kx^3)\,,
$$
and then, using the equation (\ref{expcond}) we can
find another
 Jacobi multiplier of the form
${\mathcal{D}}_2^{\nu_2}$ with $\nu_2=-4/3$.  The Abel equation
admits a Lagrangian description by means  of a function $L_2$ such
that
$$
 \fracpd{^2L_2}{v^2} = (3v+kx^3)^{-4/3}\,,
$$
from where we obtain the Lagrangian $L_2=\wt{L}_A$ given by (\ref{LagAbel2}).

Remark that, as indicated above, if $P$ and $Q$ are two Darboux
polynomials with the same cofactor then $P/Q$ is a constant of the
motion. This is just what happens with the energy $E_{L_A}$
obtained in (\ref{EL1}) which is given by
${\mathcal{D}}_2/{\mathcal{D}}_1^3$ (up to the sign).

\section{Abel equations of third and fourth-order }

In the following all the functions that appear as constants of the motion will depend on the time $t$.  So we better consider  to describe them  as first integrals that depend polynomially on time.

\subsection{Abel equation of third-order}

The action of the operator $\ID_A$ three times on the function
$x(t)$ leads to the following  nonlinear equation
\begin{equation}
\frac{d^3 x}{dt^3} + 5 k x^2 \Bigl(\frac{d^2 x}{dt^2}\Bigr) + 8 k
x \Bigl(\frac{d x}{dt}\Bigr)^2 + 9 k^2 x^4 \Bigl(\frac{d
x}{dt}\Bigr) + k^3 x^7 = 0 \,, \label{Abeleq3}
\end{equation}
that represents the third-order element of the   Abel equation
chain. It can be presented as a system of three first-order
equations
\begin{equation}
\left\{\begin{array}{rcl}
 \dfrac{dx}{d t} &=& v  \cr
\dfrac{dv}{d t} &=& a \cr \dfrac{da}{d t} &=&  - 5 k x^2 a - 8 k
x v^2 - 9 k^2 x^4v - k^3 x^7
\end{array}\right.
\end{equation}
that represents a dynamical system that, in differential geometric
terms, is represented by the following  vector field
 in the phase space $\IR^3$, with coordinates $(x,v,a)$
\begin{equation}
 \Ga^{(3)} = v\,\fracpd{}{x} + a\,\fracpd{}{v} + F_{A3}\,\fracpd{}{a}
\,,\qquad
 F_{A3} = - 5 k x^2 a - 8 k x v^2 - 9 k^2 x^4v - k^3 x^7 \,.
\end{equation}

In what follows we make use of the following polynomials
$$
  P_{A0}  = x  \,,\qquad
  P_{A1}  = v - F_{A1}  = v + k\,x^3  \,,\qquad
  P_{A2}  = a - F_{A2}  = a + 4 k x^2 v + k^2 x^5 \,,
$$
defined on the phase space and obtained by making use of the
substitution $\dot{x}\mapsto v$ and $\ddot{x}\mapsto a$. Then we
have
\begin{eqnarray*}
 \Ga^{(3)}(P_{A0}) + k x^2 P_{A0} &=& v + k\,x^3,  \cr
 \Ga^{(3)}(P_{A1}) + k x^2 P_{A1} &=&  a + 4 k x^2 v + k^2 x^5,  \cr
 \Ga^{(3)}(P_{A2}) + k x^2 P_{A2} &=& F_{A3} + 5 k x^2 a + 8 k x  v^2 + 9 k^2 x^4  v+ k^3 x^7.
\end{eqnarray*}
that can be rewritten as follows
\begin{eqnarray*}
 \Ga^{(3)}(P_{A0}) + k x^2 P_{A0} &=&  P_{A1},  \cr
 \Ga^{(3)}(P_{A1}) + k x^2 P_{A1} &=&  P_{A2},  \cr
 \Ga^{(3)}(P_{A2}) + k x^2 P_{A2} &=& 0.
\end{eqnarray*}
Note that  according to these properties $P_{A2}$ is a Darboux
polynomial with $f=-k x^2$ as cofactor. The  divergence of the
vector field  $ \Ga^{(3)}$ is $-5kx^2$, and using relation
(\ref{expcond}) we see that $R=(P_{A2})^{\mu_2}$ with $\mu_2=-5$
is a Jacobi multiplier.

 Next let $T_1^{(3)}$ be the following function
$$
  T_1^{(3)} = \frac{x}{P_{A2}}  \,,
$$
and then we have
$$
  \Ga^{(3)}(T_1^{(3)})  = T_2^{(3)} =  \frac{v + k x^3}{P_{A2}} \,,{\qquad}
  \Ga^{(3)}(T_2^{(3)})  = T_3^{(3)} = 1\,,{\qquad}
  \Ga^{(3)}(T_3^{(3)})  = T_4^{(3)} =0  \,.
 $$
This means that $T_1^{(3)}$ and $T_2^{(3)}$ are generators of of
constants of motion for the third-order element of the Abel
equation chain represented by the dynamical vector field
$\Ga^{(3)}$. Thus we can state the following proposition.

\begin{proposicion}  The two functions $J_{t1}$ and $J_{t2}$
defined as
$$
 J_{t1} = T_{2}^{(3)} - t   \,,\qquad
 J_{t2} = T_{1}^{(3)} - t \,T_2^{(3)}  + (\frac{1}{2})  t^2  \,,
$$
are first integrals, that depend polynomially on time, for the Abel equation
of third-order.
\end{proposicion}

Note that $J_{t1}$ is linear in the time $t$ and $J_{t2}$ is
quadratic.  So these expressions are similar to the constants of
the motion determined by master symmetries; nevertheless in this
third-order case we have not made use of any symplectic structure
and we have obtained these functions without relating them with
symmetries of a symplectic structure.  This is an interesting
situation deserving an additional analysis in the next sections.

Note also that both, $J_{t1}$ and $J_{t2}$, can be written as
quotients of polynomials; so if we consider the system as a
time-dependent system then the dynamics is geometrically
represented by the vector field $\Ga_t^{(3)} = \Ga^{(3)} +
\partial/\partial t$ and the following polynomials
$$
 {\mathcal{D}}_2 = P_{A1} - t P_{A2} \,,\qquad
 {\mathcal{D}}_3 = P_{A0} - t P_{A1} + (\frac{1}{2}) t^2 P_{A2} \,,
$$
are two Darboux polynomials with the same cofactor as $P_{A2}$
$$
 \Ga_t^{(3)}({\mathcal{D}}_i) = \Bigl(\Ga^{(3)}+\fracpd{}{t}\Bigr)
 ({\mathcal{D}}_i) = -\,kx^2\,{\mathcal{D}}_i \,,\quad  i=2,3.
$$

\subsection{Abel equation of  fourth-order }

The action of $\ID_A$ four times on the function $x(t)$ leads to
the following  nonlinear equation
\begin{equation}
x^{iv)} + 2 k (4 \dot{x}^3 + 13 x \dot{x}\ddot{x}+ 3 x^2 \dddot x)
 + 2 k^2 x^3 (22 \dot{x}^2 + 7 x\dot{x}) + 16k^3 x^6\dot{x} + k^4 x^9
 = 0 \,,   \label{Abeleq4}
\end{equation}
that represents the fourth-order element of the Abel equation
chain. This equation determines a dynamical system that, in
geometric terms, can be represented by the following vector field
on $\IR^4$ as phase space, with coordinates $(x,v,a,w)$:
\begin{equation}
 \Ga^{(4)} = v\,\fracpd{}{x} + a\,\fracpd{}{v} + w\,\fracpd{}{a}
 + F_{A4}\,\fracpd{}{w} \,,
\end{equation}
where
$$
 F_{A4} = -2 k (4 v^3 + 13 x v a+ 3 x^2 w)
 - 2 k^2 x^3 (22 v^2 + 7 x a) - 16k^3 x^6 v - k^4 x^9.
$$

Now we introduce the polynomial $P_{A3}$
$$
 P_{A3} = w - F_{A3} = w + 5 k x^2 a + 8 k x  v^2 + 9 k^2 x^4  v+ k^3 x^7 \,,
$$
obtained from the expression of $\ID_A^3 \,x$ with the
substitution $\dot{x}\mapsto v$,  $\ddot{x}\mapsto a$ y ${\dddot{x}}\mapsto
w$. Then we have
\begin{eqnarray*}
 \Ga^{(4)}(P_{A0}) + k x^2 P_{A0} &=& v + k\,x^3  \cr
 \Ga^{(4)}(P_{A1}) + k x^2 P_{A1} &=&  a + 4 k x^2 v + k^2 x^5  \cr
 \Ga^{(4)}(P_{A2}) + k x^2 P_{A2} &=& w + 5 k x^2 a + 8 k x  v^2 + 9 k^2 x^4  v+ k^3 x^7  \cr
 \Ga^{(4)}(P_{A3}) + k x^2 P_{A3} &=& F_{A4} + k( 8 v^3 + 26 x v a + 6 x^2 w)  + k^2 (44 x^3 v^2 + 14 x^4 a)\cr
 &&{\hskip160pt} + 16 k^3 x^6 v + k^4 x^9
 \end{eqnarray*}
that can be rewritten as follows
\begin{eqnarray*}
 \Ga^{(4)}(P_{A0}) + k x^2 P_{A0} &=&  P_{A1}  \cr
 \Ga^{(4)}(P_{A1}) + k x^2 P_{A1} &=&  P_{A2}  \cr
 \Ga^{(4)}(P_{A2}) + k x^2 P_{A2} &=&  P_{A3}  \cr
 \Ga^{(4)}(P_{A3}) + k x^2 P_{A3} &=&  0
\end{eqnarray*}

Let now  $T_1^{(4)}$ be the following function
$$
 T_1^{(4)} = \frac{x}{P_{A3}},
$$
and then we have
$$
 \Ga^{(4)}(T_1^{(4)})  = T_2^{(4)}  \,,{\qquad}
 \Ga^{(4)}(T_2^{(4)})  = T_3^{(4)}  \,,{\qquad}
 \Ga^{(4)}(T_3^{(4)})  = T_4^{(4)}   \,,{\qquad}
 \Ga^{(4)}(T_4^{(4)})  = 0    \,,
$$
with $T_2^{(4)}$, $T_3^{(4)}$, and $T_4^{(4)}$  given by
$$
 T_2^{(4)} = \frac{v + k\,x^3}{P_{A3}}   \,,\quad
 T_3^{(4)} = \frac{a + 4 k x^2 v + k^2 x^5}{P_{A3}}   \,,\quad
 T_4^{(4)} = \frac{w + \dots + k^3 x^7}{P_{A3}} = 1
$$

\begin{proposicion}
The three functions $J_{t1}$, $J_{t2}$, and $J_{t3}$ defined as
\begin{eqnarray*}
  J_{t1} &=& T_3 - t \cr
  J_{t2} &=& T_2 - t \,T_3 + (\frac{1}{2}) t^2  \cr
  J_{t3} &=& T_1 - t \,T_2 + (\frac{1}{2}) t^2 T_3 - (\frac{1}{6}) t^3
\end{eqnarray*}
are first integrals, that depend polynomially on time, for the fourth-order
element of the Abel equation chain.
\end{proposicion}

The situation is similar to the $n=3$ case and the functions
$J_{tr}$, $r=1,2,3$, are polynomials of order $r$ in the variable
$t$.

\section{Equation of Abel of order  $n$ }

We have seen that the second-order element of the Abel equation
chain  is endowed with some specific properties (e.g., it admits a
Lagrangian description) but the third and fourth-order elements
of the chain also enjoy very similar properties. Now in this
section we study the equation of order $n$ and  prove that these
properties characterise to all the equations of the family in an
independent  of the order way.

The equation of Abel of order $n$ can be obtained as the equation
arising from the action of the operator $\ID_A$ on the equation of
order $n-1$
$$
  \ID_A(\ID_A^{n-1}\,x) = \ID_A^n  x = 0  \,.
$$
This equation determines a dynamical system that, in geometric
terms, can be represented by the following vector
field defined on the phase space $\IR^n$, with coordinates $(x=x_1, x_2,
x_3,\dots, x_n)$:
\begin{equation}
  \Ga^{(n)} = x_2\,\fracpd{}{x} + x_3\,\fracpd{}{x_2} + x_4\,\fracpd{}{x_3}
  + \dots +F_{An}\,\fracpd{}{x_n},
\end{equation}
where $F_{An}$ is obtained a from the expression for $\ID_A^{n}
\,x$ with the substitution $x\mapsto x_1$, $\dot{x}\to x_2$, $\ddot{x}\to x_3$,
$\dddot{x}\to x_4$, $\dots$

In the previous sections we have made use of the polynomials
$P_{A0}$, $P_{A1}$, $P_{A2}$, and $ P_{A3}$ defined in the phase
space and whose explicit expressions, when written in the notation
of the coordinates $x_1, x_2, x_3,\dots, x_n$,  were given by
\begin{eqnarray*}
 P_{A0} &=&  x_1  \,,\cr
 P_{A1} &=&  x_2 - F_{A1} = x_2 + k\,x_1^3 \,,\cr
 P_{A2} &=&  x_3 - F_{A2} = x_3 + 4 k x^2_1 x_2 + k^2 x_1^5 \,,\cr
 P_{A3} &=&  x_4 - F_{A3} = x_4 + 5 k x^2_1 x_3 + 8 k x_1 x_2^2 + 9 k^2 x_1^4 x_2  + k^3
 x^7_1 \,.
\end{eqnarray*}
In the general case  we have $P_{An-1} = x_n -  F_{An-1}$ that
leads to an expression of the form
$$
 P_{An-1} = x_n -  F_{An-1} = x_n  + (n+1) k x^2x_{n-1} + \dots + k^{n-1} x^{2n-1}
 \,.
$$

\begin{proposicion} \label{prop4}
The action of the dynamical vector field $\Ga^{(n)}$ on the $n$
polynomials $P_{Ar}$, $r=0,1,2,\dots,n-1$, is given by
\begin{eqnarray*}
 &(i)&   {\quad} \Ga^{(n)}(P_{Ar}) + k x^2 P_{Ar} =  P_{Ar+1}
    \,,\quad r=0,1,2,\dots,n-2,   \cr
 &(ii)& {\quad} \Ga^{(n)}(P_{An-1}) + k x^2 P_{An-1} = 0.
\end{eqnarray*}
\end{proposicion}

The property  $(i)$ can be proved by induction. The property $(ii)$
follows by direct calculus.  $\Box$

\begin{proposicion} 
The time evolution of the functions $P_{Ar}/P_{An-1}$ is given by
$$
 \Ga^{(n)} \Bigl(\frac{P_{Ar}}{P_{An-1}}\Bigr)
  = \frac{P_{Ar+1}}{P_{An-1}}  \,,\quad r=0,1,2,\dots,n-2.
$$
\end{proposicion}

By direct calculus we have
$$
 \Ga^{(n)} \Bigl(\frac{P_{Ar}}{P_{An-1}}\Bigr)
  = \frac{\Ga^{(n)}(P_{Ar})P_{An-1} - P_{Ar}\Ga^{(n)}(P_{An-1})}{(P_{An-1})^2} \,.  $$
Then, making use of the properties $(i)$ and $(ii)$ of proposition (\ref{prop4}), we arrive at
$$
 \Ga^{(n)} \Bigl(\frac{P_{Ar}}{P_{An-1}}\Bigr)
  = \frac{[P_{Ar+1} - k x^2 P_{Ar}] P_{An-1}- P_{Ar}[-k x^2 P_{An-1}]}{(P_{An-1})^2}   = \frac{P_{Ar+1}}{P_{An-1}} .  {\qquad}\Box
$$

Notice that the first and the last derivatives in this series,
corresponding to $r=0$ and $r=n-2$, become
$$
 \frac{d}{dt} \Bigl(\frac{x}{P_{An-1}}\Bigr) = \frac{P_{A1}}{P_{An-1}}
 \,,\qquad
 \frac{d}{dt} \Bigl(\frac{P_{An-2}}{P_{An-1}}\Bigr) = 1  \,.
$$

Now let  $T_1^{(n)}$ be  the following function defined by
$$
  T_1^{(n)} = \frac{x}{P_{An-1}},
$$
and then,  making use of the two preceding  Propositions we can obtain the
values of the  sequence of  time derivatives of the functions
$T_{k}^{(n)}$, which are given by
\begin{eqnarray*}
  \Ga^{(n)} \bigl(T_{1}^{(n)}\bigr)&=& T_2^{(n)} = \frac{x_2 + k x^3}{P_{An-1}}
  = \frac{P_{A1}}{P_{An-1}}        \cr
  \Ga^{(n)} \bigl(T_{2}^{(n)}\bigr)&=& T_3^{(n)} = \frac{x_3 + 4 k x^2 x_2
  + k^2 x^5}{P_{An-1}}  = \frac{P_{A2}}{P_{An-1}}    \cr
  \Ga^{(n)} \bigl(T_{3}^{(n)}\bigr)&=& T_4^{(n)} = \frac{x_4 + 5 k x^2 x_3
  + 8 k x x_2^2 + 9 k^2 x^4 x_2  + k^3 x^7}{P_{An-1}}
  =  \frac{P_{A3}}{P_{Rn-1}}    \cr
  \dots &&\dots \dots \dots \dots  \dots \dots \dots\dots \dots \dots   \cr
  \dots &&\dots \dots \dots \dots  \dots \dots \dots\dots \dots \dots   \cr
  \Ga^{(n)} \bigl(T_{n-1}^{(n)}\bigr)&=& T_{n}^{(n)} = 1   \cr
  \Ga^{(n)} \bigl(T_{n}^{(n)}\bigr)   &=& 0   \,.
\end{eqnarray*}
From here we can state the existence of a family of $n-1$
first integrals depending polynomially on time.

\begin{proposicion}
The $(n-1)$ functions $J_{tr}$, $r=1,2,3,\dots,n-1$,  defined as the
following  polynomials of order $r$ in the variable $t$
\begin{eqnarray*}
 J_{t1}  &=& T_{n-1}^{(n)} - \,t   \cr
 J_{t2}  &=& T_{n-2}^{(n)} - t \,T_{n-1}^{(n)} + (\frac{1}{2}) \,t^2  \cr
 J_{t3}  &=& T_{n-3}^{(n)} - t \,T_{n-2}^{(n)} + (\frac{1}{2}) \,t^2
 \, T_{n-1}^{(n)} - (\frac{1}{6})  \,t^3        \cr
 \dots &&\dots \dots \dots \dots  \dots \dots \dots\dots \dots \dots  \cr
  J_{tn-1}  &=& T_1^{(n)} -  t \,T_2^{(n)} + (\frac{1}{2})\,t^2  \,T_3^{(n)}
  - \dots \dots +  (-1)^n (\frac{1}{n!})  \,t^n
\end{eqnarray*}
are $n-1$ functionally independent first integrals, that depend polynomially on time, for the Abel equation of order $n$.
\end{proposicion}
 
An alternative form of proving the existence of all these
constant of the motion is as follows. The $n$ polynomials
${\mathcal{D}}_a$, $a=1,2,\dots,n$, defined in the extended phase
space $\IR^n\times \IR$ as
\begin{eqnarray*}
 {\mathcal{D}}_1 &=& P_{An-1} \,,\cr
 {\mathcal{D}}_2 &=& P_{An-2} - t P_{An-1} \,,\cr
 {\mathcal{D}}_3 &=& P_{An-3} - t P_{An-2} + (\frac{1}{2}) t^2 P_{An-1}  \,,\cr
 {\mathcal{D}}_4 &=& P_{An-4} - t P_{An-3} + (\frac{1}{2}) t^2 P_{An-2}
   - (\frac{1}{6})  \,t^3 P_{An-1} \,,\cr
 \dots &&\dots \dots \dots \dots  \dots \dots \dots\dots \dots \dots\cr
 {\mathcal{D}}_{n} &=& P_{A0} - t P_{A1} + \dots +
   (-1)^n (\frac{1}{n!})  \,t^n P_{An-1} \,,
\end{eqnarray*}
are $n$ Darboux polynomials with the same cofactor
$$
 \Ga_t^{(n)}({\mathcal{D}}_a) = \Bigl(\Ga^{(n)}+\fracpd{}{t}\Bigr)
 ({\mathcal{D}}_a) = -\,kx^2\,{\mathcal{D}}_a \,,\quad   a=1,2,\dots, n.
$$
Hence the functions
$$
 J_{tab} = \frac{{\mathcal{D}}_a}{{\mathcal{D}}_b} \,,\quad
 a,b=1,2,\dots,n,
$$
are  constants of the motion. In fact, we can arrange
all these functions as the entries of an $n$-dimensional matrix
$[J_{tab}]$ that becomes a matrix formed by constants of the
motion (the diagonal elements are just ones) with the fundamental
set of functions $J_{tk}$ placed in the first row.

Finally, the  divergence of the vector field  $\Ga^{(n)}$ is given
by ${\rm div\,}\Ga^{(n)}=-(n+2)kx^2$. Thus, using relation
(\ref{expcond}), we obtain the following Jacobi multipliers for
the Abel equation of order $n$ (or for the dynamical vector field
$\Ga^{(n)}$)
$$
  R_a = ({\mathcal{D}}_a)^{\mu_n} \,,{\quad} \mu_n = -(n+2)\,,{\quad}
  a=1,2,\dots,n.
$$
We note that all these $n$ Jacobi multipliers are different, that is 
$R_b\ne R_a$,  $b\ne a$, but they are proportional by a function that is 
an integral of the motion.

\section{Final comments }

We have studied a chain  of higher-order nonlinear Abel equations
using, as starting point, the idea that they have many
similarities with the higher-order nonlinear Riccati equations. We
have made use of the Lagrangian formalism (inverse problem,
non-polynomial Lagrangians, nonstandard symmetries) in the case of
the second-order equation and of other mathematical tools (Darboux
polynomials and Jacobi multipliers) in the case of higher-order
nonlinearities. All these questions seems to be  really
interesting and we think they deserve a deeper study.

Finally, we mention that all these equations have (for any
order of the equation) a family of first integrals 
$J_{tk}$ that depend of the time as a polynomial in $t$.  In the
symplectic case functions of such a class are associated to master
symmetries of the (Lagrangian or Hamiltonian) system, but in the
general Abel case we have proved the existence of such constants
without refering to any symplectic structure. This is in fact a very
interesting fact that is to be studied.

\section*{Acknowledgments}

JFC and MFR acknowledge support from research projects
MTM-2006-10531,  FIS-2006-01225, and E24/1 (DGA). PG thanks the
Departamento de F\'{\i}sica Te\'orica de la Universidad de Zaragoza for
its hospitality and acknowledges support from Max Planck Institute
for Mathematics in the Sciences, Leipzig.


\end{document}